\documentclass[journal]{IEEEtran}

\ifCLASSINFOpdf
\else
   \usepackage[dvips]{graphicx}
\fi
\usepackage{url}

\usepackage{graphicx}

\usepackage[margin= 1 in]{geometry}

\date{\vspace{-5ex}}
\usepackage{amsthm, amsfonts, mathrsfs}
\usepackage{ dsfont }
\usepackage{amssymb}
\usepackage{enumerate}
\usepackage{graphicx}
\usepackage{float}
\usepackage{bbm}
\usepackage{amsmath}
\usepackage{comment}
\usepackage{hyperref}
\usepackage{listings}
\usepackage{multicol}
\usepackage{color}
\usepackage[dvipsnames]{xcolor}

\hypersetup{
    colorlinks,
    citecolor=blue,
    filecolor=blue,
    linkcolor=blue,
    urlcolor=blue
}

\newtheorem{theorem}{Theorem}[section]

\newtheorem{lemma}[theorem]{Lemma}

\newtheorem{definition}[theorem]{Definition}
\newtheorem{remark}[theorem]{Remark}

\newcommand{\R}{\mathbb{R}}

\newcommand{\inv}{\dagger}

\usepackage[
backend=biber,
style= numeric,
sorting=ynt
]{biblatex}
\begin{document}

\title{Signed Cumulative Distribution Transform for Parameter Estimation of 1-D Signals}

\author{Sumati Thareja, Gustavo Rohde, Rocio Diaz Martin, Ivan Medri, and Akram Aldroubi

\thanks{This work was supported in part by National Institutes of Health Grant GM130825.}
\thanks{Sumati Thareja, Rocio Diaz Martin, Ivan Medri and Akram Aldroubi are with Department of Mathematics, Vanderbilt University (e-mail: sumati.thareja, rocio.p.diaz.martin, ivan.v.medri, akram.aldroubi@vanderbilt.edu).}
\thanks{Gustavo Rohde is with Department of Biomedical Engineering, Department of Electrical and Computer Engineering, University of Virginia (email: gr2z@virginia.edu).}}

\maketitle

\begin{abstract}
We describe a method for signal parameter estimation using the signed cumulative distribution transform (SCDT), a recently introduced signal representation tool based on optimal transport theory. The method builds upon signal estimation using the cumulative distribution transform (CDT) originally introduced for positive distributions. Specifically, we show that Wasserstein-type distance minimization can be performed simply using linear least squares techniques in SCDT space for arbitrary signal classes, thus providing a global minimizer for the estimation problem even when the underlying signal is a nonlinear function of the unknown parameters. Comparisons to current signal estimation methods using $L_p$ minimization shows the advantage of the method.
\end{abstract}



\section{Introduction}

\begin{figure}
        \centerline{\includegraphics[width=\columnwidth]{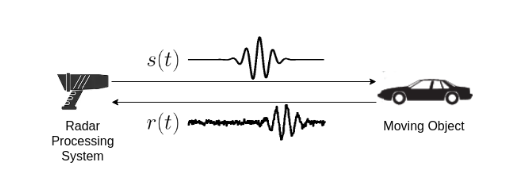}}
        \caption{Estimation in Radar signal processing applications. In radar systems, the estimated time delay and Doppler stretch between transmitted and received signals are used to determine the position and speed of a target object. (Image credits \cite{rubaiyat2020parametric}).}
        \label{Radar}
    \end{figure}

 \IEEEPARstart{S}{everal} problems in science and engineering require one to estimate certain parameters $\textbf{p}=\{p_n\}_{n=0}^k$, 
 of a function $g_\textbf{p}(t)$ (e.g., $g_\textbf{p}(t) = \sum\limits_{n=0}^{k}p_{n}t^n$) that produces a signal via the generative model
    \begin{equation} \label{PDF_Composition}
        s_{g_{\textbf{p}}}(t) = g'_{\textbf{p}}(t)s(g_{\textbf{p}}(t))
     \end{equation}
    that best matches some measured signal $r(t)$. In other words, we seek $\textbf{p}$ such that $s_{g_{\textbf{p}}}(t) \sim r(t)$, $t \in \Omega$ with $\Omega$ some measured interval of time. Example applications include time delay ($g_\tau(t) = t-\tau$), velocity ($g_{\tau,b}(t) = bt-\tau$) or acceleration ($f_{\tau,a,b}(t) = at^2 +  bt-\tau$) estimation problems \cite{rubaiyat2020parametric} (see Fig. \ref{Radar}), source localization problems \cite{Amar:12, Nichols:19}, communications \cite{Proakis:83}, among others. As far as motion is concerned, it is common to utilize position, velocity, acceleration, jerk, etc. as \textit{parametric} descriptions of a particle system, and thus polynomials  $g_\textbf{p}(t) = \sum\limits_{n=0}^{k}p_{n}t^n$ are natural choices and represent an important example to work on. Other linear models such as $g_\textbf{p}(t) = \sum\limits_{n=0}^{k}p_{n}\phi_n(t)$ are also possible, including B-splines, RBFs, wavelets, Fourier basis, or vector learned via principal component analysis techniques, for example.

\begin{figure}
\centerline{\includegraphics[scale=0.5,width=\columnwidth]{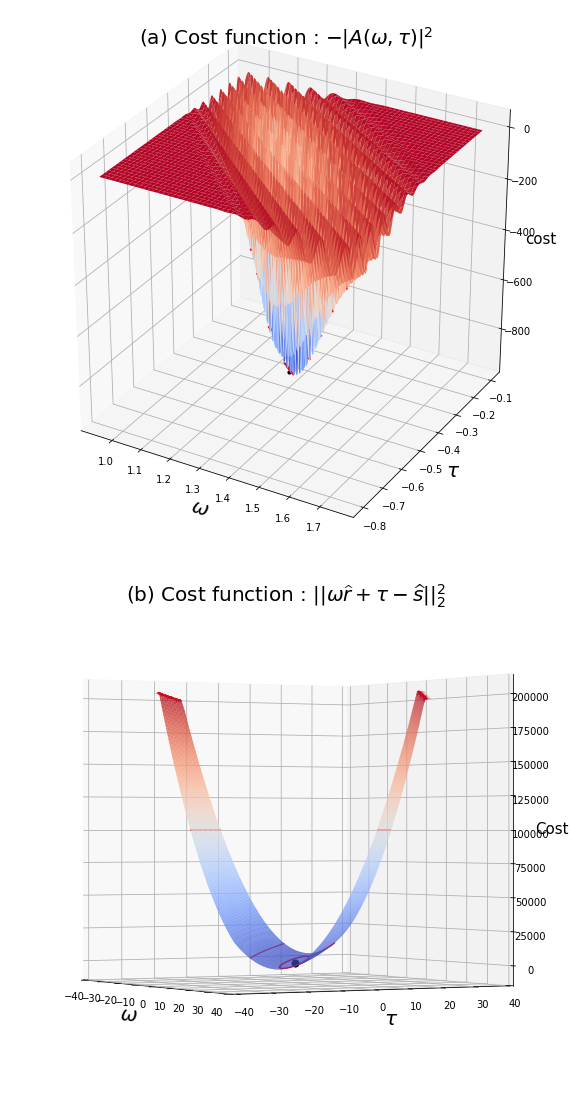}}
\caption{Cost functions associated with joint time delay and linear dispersion estimation for the original signal $s(t)$ and received signal $r(t)$ (a) using Wide-band Ambiguity Function (WBAF) where $A(\omega,\tau)=\sqrt{\omega}\int_\mathbb{R} r(t) \, s(\omega t+\tau) dt$ and (b) using proposed SCDT based estimator (black dot shows the global minimum point).}
\label{Cost3d}
\end{figure}

     In the presence of Gaussian distributed noise, following the maximum likelihood principle, the least squares metric is often used to measure the agreement between $s_{g_{\textbf{p}}}(t)$ and $r(t)$. When $g_{\textbf{p}}(t) = t-p_0$, Fourier transform techniques can be used to accelerate the least squares solution of 
    \begin{equation} \label{old_esti}
        \min_{p_0} \| s(t-p_0) - r(t) \|^2_2
    \end{equation}
    by maximizing the cross correlation in Fourier transform space \cite{jacovitti93}. Likewise, Fourier transform techniques can also be used to find the frequency and phase shifts of a model signal in such a way that it will best match an input signal $r(t)$. However, estimating the parameters of $s_{g_{\textbf{p}}}(t)$, when $g_{\textbf{p}}(t)$ goes beyond a linear polynomial, is not as straight forward, and usually requires nonlinear, non convex (see Fig. \ref{Cost3d}(a)), global optimization approaches \cite{Jin95}, \cite{Niu99},  \cite{Tao2008TwostageMF}, \cite{Colonnese2010}.

The approach that we take is to change the Euclidean distance used above in  \eqref{old_esti}, 
into a new, Wasserstein-type, metric for measuring the agreement between two signals. For that, we use the $L^2$ metric in the space of functions by means of SCDT, a new signal transform recently introduced in \cite{aldroubi2021signed}, that converts the generative model into a simpler expression. In this manner we are able to change a non-convex optimization problem in the native space into a linear, convex problem (see Fig. \ref{Cost3d}(b)) in the SCDT transform space, by minimising the functional,
    
    \begin{align} \label{functional}
        J(\textbf{p}) := \, & d_{W^2}^2\left(\frac{r^+}{\|r^+\|_1},\frac{{s^{+}_{g_{\textbf{p}}}}}{\|{s^{+}_{g_{\textbf{p}}}}\|_1}\right)  \notag\\
        &  +  d_{W^2}^2\left(\frac{r^-}{\|r^-\|_1},\frac{{s^{-}_{g_{\textbf{p}}}}}{\|{s^{-}_{g_{\textbf{p}}}}\|_1}\right)  \,
    \end{align}
    over all $\textbf{p}.$ Here, $r^{+}=\max\{0,r\}$ and $r^{-}=-\min\{0,r\}$ (resp. ${s^{\pm}_{g_{\textbf{p}}}}$) are the positive and negative parts of $r$ (resp. of ${s_{g_{\textbf{p}}}}$) and $d_{W^2}(\cdot, \cdot)$ is the usual Wasserstein distance \cite{villani2003topics} defined for probability densities.\par
    
    The motivation for the functional $J$ comes from an attempt to generalize the Wasserstein distance between two functions $r$ and $s_{g_{\textbf{p}}}$ that are not necessarily probability density functions. One of the main advantage of this particular generalization is that it renders the parameter estimation problem as a linear problem.
    This will be shown in more detail once we introduce the SCDT transform and its properties.
    
    In section \ref{esti_solution}, we will see how the SCDT can be used to simplify the solution of \eqref{functional}.

\section{The Signed Cumulative Distribution Transform} \label{SCDT}
This section is a brief description of the SCDT (for details, see \cite{aldroubi2021signed}). Let $s$ be a non-negative signal with $\|s\|_1=1.$ The cumulation $F_s$ of $s$,  is defined as 
\begin {equation} \label {cumulation}
F_s(x) :=\int\limits_{-\infty}^x s(t)dt,
\end{equation} 
For a fixed, non-negative, normalized reference signal $s_0$, the Cumulative Distribution Transform $\mathcal {C}(s)$ of $s$ with respect to $s_0$, defined in \cite{park2017cumulative},  can be written as (see \cite{aldroubi2021signed}) 
\begin{equation} \label{CDT_probs}
        \mathcal {C} (s):={F_s}^{\inv}\circ {F_{s_0}}, 
    \end{equation}
where ${F_s}^{\inv}$ is the generalized inverse of $F_s$ defined by 
\begin{equation*}
        F^{\inv}(y) := \inf\{x: F(x)>y\}.
    \end{equation*}
For a signed signal $s,$ the transform is defined by utilising the Jordan decomposition of $s$, namely
$$s^+(x)=\max\{0,s(x)\} ,\hspace{0.1cm} s^-(x)=\max\{0,-s(x)\},$$
\noindent
then the SCDT of $s$ with respect to a non-negative, normalized reference $s_0$ is defined as,
    \begin{equation}\label{signed}
         \widehat{s} = \left((s^{+})^\star, \|s^+\|_1, (s^{-})^\star, \|s^-\|_1\right),
    \end{equation}
    where
    \begin{equation*}
    (s^{+})^\star = \begin{cases}\mathcal {C}\big(\frac{s^+}{\|s^+\|_1}\big) & \text{ if }  s^{+} \text{ is non-trivial}\\
    0 & \text{ if } s^{+}=0
    \end{cases}  
\end{equation*}
where  the operator $\mathcal {C}$ is defined in \eqref {CDT_probs}. The function $(s^{-})^\star$ is defined analogously.

    \begin{subsection} {Properties of SCDT}
    \end{subsection}
    
    The following two lemmas together will reduce the estimation problem to a linear least squares problem (for proofs see \cite{aldroubi2021signed}):

\begin{lemma} (Composition Property) \label{Composition_Signed} \cite{aldroubi2021signed}
     Let $s \in L^{1}(\R)$, and let $g : \R \rightarrow \R$ be a strictly increasing surjection. Consider $s_g: \R \rightarrow \R$ given by 
    $s_g(x) = g'(x) \, s \circ g(x) .$ Then
    $\|s_g^{\pm}\|_1 = \|s^{\pm}\|_1$, and the SCDT of $s_g$ is given by
    $$\widehat{s}_g = (g^{-1}\circ (s^{+})^{\star}, \| s^+\|_1, g^{-1}\circ (s^{-})^{\star}, \|s^-\|_1).$$
\end{lemma}


In order to state the second key lemma, we require a metric on the native space. This metric is a generalization of Wasserstein distance \cite{villani2003topics} defined for probability densities, to non-normalized signed signals.

\begin{definition} \label{distance}
Let $r, s\in L^1(\R)$ such that $\int r(x) \, |x|^2 \, dx<\infty$ and $\int s(x) \, |x|^2 \, dx<\infty$ , then
    \begin{multline} \label{gen_Wasser}
    D^2_S(r,s) := d^{2}_{W^2}\left(\frac{r^+}{\|{r}^+\|_1},\frac{s^+}{\|s^+\|_1}\right) + |\|{r}^+\|_1 -  \|s^+\|_1|^2
 \\+ d^{2}_{W^2}\left(\frac{{r}^-}{\|{r}^-\|_1},\frac{s^-}{\|s^-\|_1}\right) + |\|{r}^-\|_1 - \|s^-\|_1|^2 
    \end{multline}

\end{definition}

Utilising the SCDT and the metric $D_S(\cdot,\cdot)$, the following result is a generalization of a well-known isometry,

\begin{lemma} (Isometery) \label{isometry} \cite{aldroubi2021signed}
    Let $s,r\in L^1(\R)$ such that $\int s(x) \, |x|^2 \, dx<\infty$ and $\int r(x) \, |x|^2 \, dx<\infty$ then 
    \begin{align*}\label{wass_prob2}
         D^{2}_{S}(s,r) &= \|\widehat{s} -\widehat{r} \|^{2}_{(L^2(s_0(x)dx)\times \R)^2}  \\
         &= \|\widehat{s^+} -\widehat{r^+} \|^{2}_{L^2} + | \|s^+\| -\|r^+\| |^2 \\
         &+ \|\widehat{s^-} -\widehat{r^-} \|^{2}_{L^2} + |\|s^-\| -\|r^-\| |^2
    \end{align*}
where $\|\cdot\|_{L^2(s_0(x)dx)}$ is the norm defined by
$$\|f\|_{L^2(s_0(x)dx)}=\left(\int |f(x)|^2 \, s_0(x) \, dx \right)^{\frac{1}{2}}.$$
\end{lemma}
\noindent For notational simplicity, we will write $\|\cdot\|$ instead of $\|\cdot\|_{(L^2(s_0(x)dx)\times \R)^2}$.

\begin{remark} \label{mass_off}
Without loss of generality, we ignore the $L^1$ norm coordinates (i.e. coordinate 2 and 4) of the transform (see \eqref{signed}) for the minimisation functional (see \eqref{functional}). 
\end{remark}

\subsection{Numerical Implementation of SCDT} \label{Num}

In \cite{aldroubi2021signed} the SCDT is defined for continuous time signals. Here, we describe the numerical method for approximating the SCDT given discrete data. Let $\textbf{s} = [s_1, s_2, \cdots s_N]^{T}$ be a $N-$point
discrete time signal, where $\textbf{s}[n] = s_n$ for all $n = 1,2,\cdots N$ is the $n$th sample of $\textbf{s}$. For a particular choice of reference signal $\textbf{s}_0$ (e.g., $\textbf{s}_0 = [1,1,\cdots 1]$), the SCDT requires computation of the positive ($\textbf{s}^+$) and the negative ($\textbf{s}^-$) part of $\textbf{s}$, their $\ell^1 -$ norms and then the corresponding cumulation functions (see \eqref{cumulation}).

Now, $\textbf{s}^+ = \frac{\textbf{s} + |\textbf{s}|}{2}$ and $\textbf{s}^- = \frac{\textbf{s} - |\textbf{s}|}{2}$, where these algebraic operations are coordinate-wise.
Also, in the discrete case, $\|\textbf{s}\|_1 = \displaystyle \sum_{k=1}^{N} |s_k|$ for any signal $\textbf{s}.$
Finally, for a signal $\textbf{s}$ the numerical approximation of the cumulation $F_{\textbf{s}}$ is given by, $F_{\textbf{s}}[n] = \displaystyle \sum_{k=1}^{n} s_k.$ The SCDT is then calculated by, taking the generalized inverse of the cumulation of $\textbf{s}$ evaluated at each coordinate of cumulation $F_{\textbf{s}_0}$ of the reference signal, using discrete generalized inverse we have implemented in \cite{Codes3}. 

\section{Solving the Estimation Problem} \label{esti_solution}

This section will demonstrate the use of SCDT in estimating signal parameters described in \eqref{PDF_Composition}. Using Lemma \ref{Composition_Signed} and Lemma \ref{isometry}, the cost function \eqref{functional} becomes,
\begin{equation} 
D_S(r,s_{g_\textbf{p}}) = \|\widehat{r} -\widehat{s}_{g_\textbf{p}} \| = \|\widehat{r} - g^{-1}_{\textbf{p}} \circ \widehat{s} \|
\end{equation}
where $r$ is the measured signal. 

When $g^{-1}_{\textbf{p}},$ does not have a closed form, we use an equivalent formulation of the minimisation problem, where we seek a match between $r_{f_{\textbf{q}}}$ and $s,$ where $r_{f_{\textbf{q}}}(t) = f'_{\textbf{q}}(t) r(f_{\textbf{q}}(t))$ and $f_{\textbf{q}} = g^{-1}_{\textbf{p}}.$ The minimisation problem then becomes,
\begin{equation} 
    D_S(r_{f_{\textbf{q}}},s) = \|\widehat{r}_{f_{\textbf{q}}}-\widehat{s} \| = \|f^{-1}_{\textbf{q}} \circ \widehat{r} -  \widehat{s} \|.
\end{equation}
Therefore we get,
\begin{equation} \label{esti_prob}
    D_S(r_{f_{\textbf{q}}},s) = \|f^{-1}_{\textbf{q}} \circ \widehat{r} -  \widehat{s} \| = \|g_{\textbf{p}} \circ \widehat{r} -  \widehat{s} \|.
\end{equation}
As shown below, if $g_\textbf{p}$ is a polynomial of degree $k$, or indeed any linear model, then the estimation problem in the SCDT domain becomes a linear least squares problem.

\subsection{Polynomial Estimation}
Consider a polynomial $g(t) = \displaystyle \sum_{k=0}^{n} p_k t^k$, then  $$g \circ \widehat{r} = \displaystyle \sum_{k=0}^{n} p_k (\widehat{r})^k,$$ therefore, the cost function (\ref{esti_prob}) then becomes 
    $D^2_S(r_f,s) = \|\sum_{k=0}^n p_k(\widehat{r})^k - \widehat{s}\|^2.$
Using the discretization technique described in section \ref{Num}, and utilizing the same symbol for a function and its vector dicretization, we get the discrete approximation $\mathbb D_{S}$ of  $D_S$:
\begin{align*}
        \mathbb D_S(r_f,s) 
        & = \left| \left|
            \begin{bmatrix} | & | & & |\\ 1 & \widehat{r} &\cdots & \widehat{r}^n \\ | & | & & | \end{bmatrix}   \begin{bmatrix} |\\ \textbf{p} \\ | \end{bmatrix} - \begin{bmatrix} |\\   \widehat{s} \\ | \end{bmatrix} \right|\right|
\end{align*}
where $\textbf{p} = [p_0,p_1, \cdots, p_n]^T.$

Using lemma \ref{isometry} and remark \ref{mass_off} the estimation problem to be solved is now reduced to solving the linear least squares problem, in the transform domain,
\begin{equation} \label{poly_sol}
    \widetilde{\textbf{p}} = \displaystyle argmin_{\textbf{p}} \left(\|\textbf{X}^+\textbf{p} - \widehat{\textbf{s}^+}\|^2 + \|\textbf{X}^-\textbf{p} - \widehat{\textbf{s}^-}\|^2\right)
\end{equation}
where $ \textbf{X}^{\pm} = [\textbf{1},\widehat{\textbf{r}^{\pm}},(\widehat{\textbf{r}^{\pm}})^2,\cdot,\cdot,\cdot, (\widehat{\textbf{r}^{\pm}})^n].$ The discrete estimation problem  \eqref{poly_sol} is convex. Moreover, since the matrices $\textbf{X}^{\pm}$ are of Vandermonde type, their columns are linearly independent as long as $r$ has at least $N\ge n$ distinct values.  Under this condition, the Hessian $\textbf{X}^T\textbf{X}$ is invertible. Hence, if $r$ has at least $N\ge n$ distinct values, the estimation problem in equation (\ref{poly_sol}) posseses a closed form solution given by
$\widetilde{\textbf{p}} = (\textbf{X}^T\textbf{X})^{-1}\textbf{X}^T\widehat{\textbf{s}},$
where, $\textbf{X} = [\textbf{X}^+, \textbf{X}^-]^T$ and $\widehat{\textbf{s}} = [\widehat{\textbf{s}^+},\widehat{\textbf{s}^-}]^T.$

\subsection{Time Delay and Linear Dispersion Estimation}
If $g(t) = \omega t + \tau,$ we get, $g \circ \widehat{r} = \omega \widehat{r} + \tau,$ and the cost function becomes,
$D^2_S(r_f, s) = \|\omega \widehat{r^+} + \tau - \widehat{s^+}\|^2 + \|\omega \widehat{r^-} + \tau - \widehat{s^-}\|^2$. The closed form solution to this problem is given by $[\omega, \tau]^{T} = (\textbf{X}^T\textbf{X})^{-1}\textbf{X}^T\widehat{\textbf{s}}$ where $\textbf{X} = [\widehat{\textbf{r}^\pm}, \textbf{1}]$ and $\widehat{\textbf{s}} = [\widehat{\textbf{s}^+},\widehat{\textbf{s}^-}]^T.$

\subsection{Quadratic Dispersion with Time Delay}
If $g(t) = \kappa t^2 + \tau,$ we get $g \circ \widehat{r} = \kappa \widehat{r}^2 + \tau,$ and the cost function becomes,
$D^2_S(r_f, s) = \|\kappa (\widehat{r^+})^2 + \tau - \widehat{s^+}\|^2 + \|\kappa (\widehat{r^-})^2 + \tau - \widehat{s^-}\|^2$. The closed form solution to this problem is given by $[\kappa, \tau]^{T} = (\textbf{X}^T\textbf{X})^{-1}\textbf{X}^T\widehat{\textbf{s}}$ where $\textbf{X} = [(\widehat{\textbf{r}^\pm})^2, \textbf{1}]$ and $\widehat{\textbf{s}} = [\widehat{\textbf{s}^+},\widehat{\textbf{s}^-}]^T.$

\subsection{Experiments} 
\begin{figure} 
\centerline{\includegraphics[width=\columnwidth]{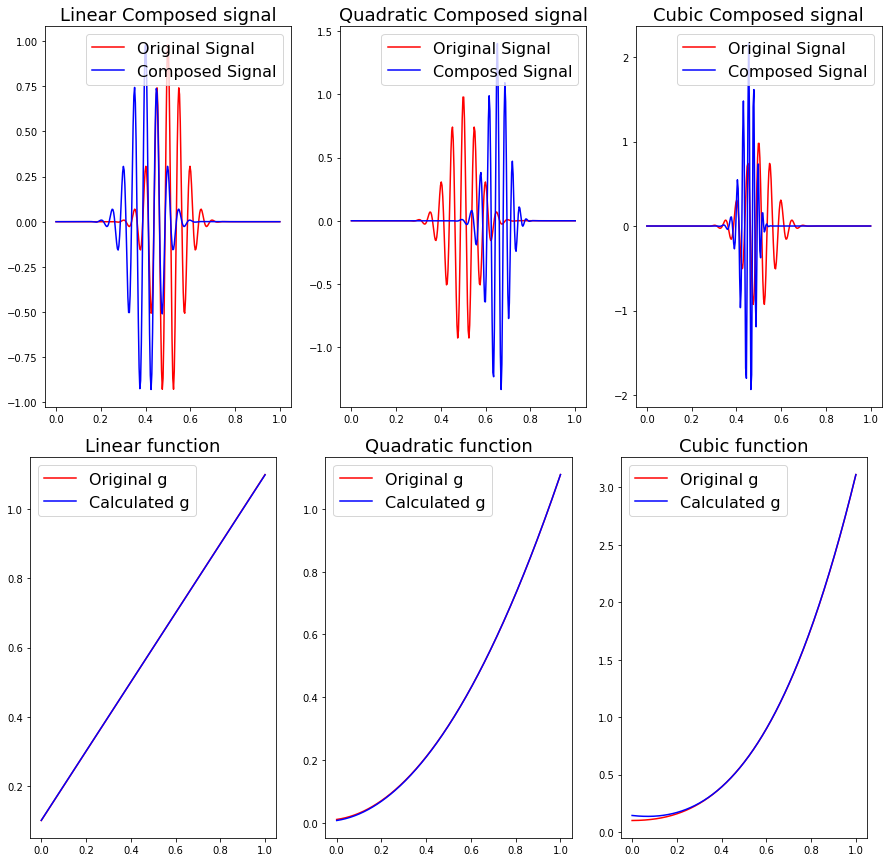}}
\caption{(Top) Signals for estimation of parameters and (bottom) Graphs of original and calculated parameter function $g(t)$ with (from left to right) linear, quadratic and cubic composition functions respectively, onto the original Gabor signal. }
\label{Estimation}
\end{figure}

To illustrate the technique, we considered the Gabor signal subject to a linear $g(t) = t + 0.1,$ a quadratic $g(t) = t^2 + 0.1 t + 0.01$ and a cubic $g(t) = 2t^3 + t^2 + 0.01 t + 0.1$ transformations, one at a time. The input signal $s(t) = \cos(40\pi(t-0.5)) \exp(-120((t-0.5)^2)),$ becomes $s_{g}(t) = 0.1 \cos(40\pi(0.1 t - 0.49)) \exp(-120((0.1 t - 0.49)^2)) $ when $g$ is linear, similarly when $g$ is quadratic and cubic, as shown in Fig. \ref{Estimation} (top). For this experiment the corresponding cumulations and the SCDT of the signals, are determined numerically as described in section \ref{Num}, using the Python code \cite{Codes3}. As can be seen in Fig. \ref{Estimation} (bottom), there is a good match between the calculated function and the function $g$ for each of the three polynomials. 

Besides the estimation problem for the polynomials as described above, 
other estimation problems can also be solved explicitly. For example, the case when $g_\textbf{p} (t) = e^{at+b}, g^{-1}_{\textbf{p}} \circ \widehat{s} = \frac{1}{a}\ln(\widehat{s}) - \frac{b}{a}, $ or when $g_\textbf{p} (t) = \ln(at+b)$ with $a\neq 0$ then $ g^{-1}_{\textbf{p}} \circ \widehat{s} = \frac{1}{a}\exp({\widehat{s}}) - \frac{b}{a}.$

\section{Conclusion}

In this paper, we proposed a parametric signal estimation approach by minimizing a Wasserstein-type distance between measured and model signals. This approach, aided by the use of the signed cumulative distribution transform, was shown to produce generic closed form solution to the estimation problem. The technique transforms a non-linear and non-convex problem in native domain into a convex, linear least square problem in the transform domain. In addition, when the the parameters to be estimated are the coefficients of a polynomial, the optimization problem has an explicit solution in the transform domain. Unlike the  technique described in \cite{rubaiyat2020parametric}, the one developed in this note can be applied to all signals that can be described by signed measures. In particular, the methods can be applied to any estimation problem carried by finite energy signals. However, currently, this method does not work effectively in the presence of noise. In order to adapt the method for signals with additive noise, there is a need of a thorough study of how the noise behaves in the transform domain.  This task is not easy because of the non-linearity of the SCDT, and will be the subject of future work.

In short, by using the SCDT and the Wasserstein cost, one can easily and accurately estimate the parameters that govern the modification of signal energy during propagation through an ideal (noiseless) medium. However, the propagation mediums are almost never ideal, adding noise to the measurements, and the effect of additive noise must be tackled by studying its characteristics after going through the SCDT, and by finding methods to minimize its effects on the estimation problem.



\printbibliography

\end{document}